\documentclass[preprint,preprintnumbers]{revtex4}

\usepackage{amssymb}
\usepackage{amsmath}
\usepackage{graphicx}
\usepackage{booktabs}
\usepackage{tabularx}
\usepackage{dcolumn}
\usepackage{bm}
\usepackage{color}
\usepackage{float}

\begin{document}

\title{Controlling Symmetries and Quantum Criticality in the Anisotropic Coupled-Top Model}
\author{Wen-Jian Mao $^{1}$, Tian Ye $^{1,2}$, Liwei Duan $^{3,}$* and Yan-Zhi Wang $^{1,2,}$*}
\address{
$^{1}$ School of Physics and Electronic Information, Anhui Normal University, Wuhu 241002, China \\
$^{2}$ Anhui Province Key Laboratory for Control and Applications of Optoelectronic Information Materials, Anhui Normal University, Wuhu 241002, China  \\
$^{3}$ Department of Physics, Zhejiang Normal University, Jinhua 321004, China
 }\date{\today }

\begin{abstract}
We investigate the anisotropic coupled-top model, which describes the interactions between two large spins along both $x-$ and $y-$directions. By tuning anisotropic coupling strengths along distinct directions, we can manipulate the system's symmetry, inducing either discrete $Z_2$ or continuous U(1) symmetry. In the thermodynamic limit, the mean-field phase diagram is divided into five phases: the disordered paramagnetic phase, the ordered ferromagnetic or antiferromagnetic phases with symmetry breaking along either $x-$ or $y-$direction. This results in a double degeneracy of the spin projections along the principal direction for $Z_2$ symmetry breaking. When U(1) symmetry is broken, infinite degeneracy associated with the Goldstone mode emerges. Beyond the mean-field ansatz, at the critical points, the energy gap closes, and both quantum fluctuations and entanglement entropy diverge, signaling the onset of second-order quantum phase transitions. These critical behaviors consistently support the universality class of $Z_2$ symmetry. Contrarily, when U(1) symmetry is broken, the energy gap vanishes beyond the critical points, yielding a novel exponent of 1, rather than 1/2 for $Z_2$ symmetry breaking. The framework provides an ideal platform for experimentally controlling symmetries and investigating associated physical phenomena.
\end{abstract}

\maketitle

\section{Introduction}

Quantum phase transitions have been a prominent area of research in various interacting systems, including fermionic \citep{Fermionic}, bosonic \citep{Bosonic} and spin systems \citep{Spin}. Unlike classical phase transitions driven by thermal fluctuations, quantum phase transitions occur at zero temperature due to symmetry breaking \citep{Spin,Symmetry} and quantum fluctuations. In spin interaction systems, numerous prototype models, such as the transverse-field Ising model (TFIM) and its generalized models \citep{TFIM1,TFIM2,TFIM3,TFIM4,TFIM5,TFIM6,TFIM7,TFIM8}, have been extensively studied to explore quantum phase transitions in both the ground states and dynamics. Recently, several approaches involving anisotropic coupling have attracted attention for investigating phase transition phenomena \citep{ATFIM1,ATFIM2,ATFIM3,ATFIM4}. Specifically, the anisotropic XY model \citep{ATFIM1}, the anisotropic two-dimensional Ising case with triangular or rectangular lattice \citep{ATFIM2,ATFIM3}, and the anisotropic next-nearest-neighbor Ising model \citep{ATFIM4}, provide insights into the system’s symmetries and critical properties.

When the the $J=1/2$ spins in the TFIM are replaced with large spins, the coupled-top model (CTM) is generalized \citep{CT1,CT2,CT3,CT4,KCT2}. As a paradigmatic bipartite system, the CTM describes the interaction between two large spins. Continuous investigations of dynamical and statistical behaviors, including ergodic behavior, quantum scars and level spacing distribution \cite{CT2,CT3,CT4}, have been conducted. In particular, various types of quantum phase transitions, such as the dynamical, ground-state and excited-state phase transitions in the CTM and its generalized cases \citep{DP1,DP2,DP3,CT4,EQPT}, have attracted significant attention. The detection of novel physical phenomena can be effectively carried out using the "out-of-time-order correlator" approach (OTOC) \citep{OTOC1,OTOC2}, which is experimentally implementable \citep{OTOC3,OTOC4}. Recently, the triangular coupled-top model has emerged as a new platform for studying unconventional frustrated magnetic behaviors, exhibiting phenomena that warrant further investigation \citep{TriCT}.

Unlike the TFIM, which is typically studied on a one-dimensional spin chain or higher-dimensional lattices, the coupled-top model is a bipartite system. This model is more analogous to the finite-component light-matter interaction systems \citep{Rabi1,Rabi2,Dicke1}, undergoing the second-order quantum phase transitions from the disorder to ordered phase \citep{Rabi3,Dicke2,SBM1,SBM2}. Paradigmatic models, such as Rabi, Dicke and spin-boson model, allow control over symmetries either $Z_2$ or U(1) by tuning anisotropic coupling strengths along two orthogonal directions \citep{ARabi1,ARabi2,ARabi3,ADicke,ASBM}. Notably, in systems with either symmetry preservation or breaking, anisotropic matter-light interactions induce novel phase diagrams and quantum triple points \citep{ARabi3,ADicke,ASBM}, which have received widespread attention. Significantly, these models reduce to the Jaynes-Cummings model \citep{JC}, Tavis-Cummings model \citep{TC} and spin-boson model \citep{RWA} with rotating wave approximation when $Z_2$ symmetry transitions to U(1) symmetry, which exhibit distinct critical behaviors related to the Goldstone mode \citep{Gold1,Gold2}. Similarly, the anisotropic coupled-top model (ACTM) can be constructed by adjusting coupling constants along $x-$ and $y-$ direction anisotropically. This scheme provides an important platform for both experimental and theoretical control of symmetries, as well as for investigating associated physical properties.

In this paper, we investigate quantum phase transitions of the anisotropic coupled-top model resulting from $Z_2$ and U(1) symmetry breaking and the associated quantum criticality. The paper is organized as follows. In Section \ref{S2}, we define the anisotropic coupled-top model and explain the $Z_2$ and U(1) symmetries. In Section \ref{S3}, we derive the mean-field solutions and corresponding phase diagram. In Section \ref{S4}, we extend the analysis beyond the mean-field approach, calculating the excitation energy, quantum fluctuations, entanglement entropy, and critical phenomena. In Section \ref{S5}, we examine the energy gap related to the Goldstone mode for U(1) symmetry. Finally, Section \ref{S6} provides the conclusions of this study.

\section{The Model and Symmetries}\label{S2}

As a generalized transverse-field Ising model (TFIM) with anisotropic couplings, the anisotropic coupled-top model (ACTM) describes interactions between two large spins along both $x$- and $y$-directions, the Hamiltonian of which is determined by
\begin{eqnarray}
\hat{H}_{\text{ACTM}}=-\epsilon \left( \hat{J}_{1}^z + \hat{J}_{2}^z \right)+ \frac{1}{J} \left( \chi_x \hat{J}_{1}^x \hat{J}_{2}^x + \chi_y \hat{J}_{1}^y \hat{J}_{2}^y \right)
\label{eq1}
\end{eqnarray}
where $\hat{J}_{i}^{d}$ denotes the $d$-component $(d=x, y, z)$ collective spin operators of magnitude $J$ for the $i$-th large spin. The parameter \(\epsilon\) characterizes the strength of an external field along the $z$-axis, which stabilizes the paramagnetic phase (PP). \(\chi_x\) and \(\chi_y\) represent anisotropic coupling constants between spins in the $x$- and $y$-directions, respectively. In particular, positive values of $\chi_x$ and $\chi_y$ drive the emergence of the antiferromagnetic phase (AFP), whereas negative values give rise to the ferromagnetic phase (FP). The competition between the three quantum phases solely depends on the parameters mentioned above. For clarity, we express the dimensionless Hamiltonian as
\begin{eqnarray}
\hat{H} = \frac{\hat{H}_{\text{ACTM}}}{\epsilon} =
 -\left( \hat{J}_{1}^z + \hat{J}_{2}^z \right)
+ \frac{1}{J} \left( \lambda_x \hat{J}_{1}^x \hat{J}_{2}^x + \lambda_y \hat{J}_{1}^y \hat{J}_{2}^y \right).
\label{eq2}
\end{eqnarray}
with dimensionless constants \(\lambda_x=\chi_x/\epsilon\) and  \(\lambda_y=\chi_y/\epsilon\).

Significantly, by tuning coupling constants $\lambda_x$ and $\lambda_y$, the anisotropic coupling system bridges the gap between discrete $Z_2$ symmetry and continuous U(1) symmetry. In the case of $|\lambda_x|\neq|\lambda_y|$, the system processes $Z_2$ symmetry, analogous to behaviors observed in Rabi or Dicke model, whereas for $|\lambda_x|=|\lambda_y|$, it displays U(1) symmetry similar to Jaynes-Cummings or Tavis-Cummings model. Therefore, this implies that by adjusting $\lambda_x$ and $\lambda_y$, we can control the presence and breaking of these symmetries, leading to distinct quantum phase transitions and corresponding physical properties.

$Z_2$ symmetry in the anisotropic system manifests the invariance of  Hamiltonian under the spin flipping along x- and y-axis, which is performed as $\hat{J}_{i}^x \to -\hat{J}_{i}^x$, $\hat{J}_{i}^y \to -\hat{J}_{i}^y$ and $J_i^z$ in the original form. Accordingly, Eq.(\ref{eq2}) satisfies the commutation relation $[\hat{H},\hat{\Pi}]=0$ with the parity operator
\begin{eqnarray}
\hat{\Pi}=\exp\left[i\pi\sum_{i=1,2}\left(\hat{J}_i^z+J\right)\right]
\label{eq3}
\end{eqnarray}
When the $Z_2$ symmetry is conserved, the system satisfies $\left\langle\hat{J}_i^x\right\rangle=0$ and $\left\langle\hat{J}_i^y\right\rangle=0$ corresponding to the disordered paramagnetic phase (PP). In contrast, when the symmetry is broken in $x-$ or $ y-$direction, $\left\langle\hat{J}_i^x\right\rangle\neq0$ or $\left\langle\hat{J}_i^y\right\rangle\neq0$ characterizes the ordered antiferromagnetic phase (AFP) and ferromagnetic phase (FP). Additionally, in the regime where $|\lambda_x|>|\lambda_y|$, the $Z_2$ symmetry breaking tend to occur in the $x$-direction, where the FP (AFP) exhibits doubly degenerate behaviors $\left\langle\hat{J}_1^x\right\rangle=\pm\left\langle\hat{J}_2^x\right\rangle$ for negative (positive) values of $\lambda_x$. Conversely, in the case of $|\lambda_x|<|\lambda_y|$, symmetry breaking along the $y$-direction predominates with analogous properties. This indicates that $\left\langle\hat{J}_i^d\right\rangle$ $(d=x,y)$ can serve as order parameters.

U(1) symmetry takes place when \(|\lambda_x|=|\lambda_{y}|\), where the Eq.(\ref{eq2}) becomes symmetric under continuous rotations in the
$x-y$ plane with angular $\varphi_i$.  Specifically, the transformation can be described as
\begin{align}
\hat{J}_i^+ \to \hat{J}_i^+ e^{i\varphi_i}, \quad \hat{J}_i^- \to \hat{J}_i^- e^{-i\varphi_i}, \quad \hat{J}_i^z=\hat{J}_i^z.
\end{align}
with $\varphi_1=\pm\varphi_2$ corresponding to $\lambda_x=\pm\lambda_y$ respectively. When U(1) symmetry is broken, the ground states lie at boundaries between FP and AFP, which satisfies $\left\langle\hat{J}_i^x\right\rangle^2+\left\langle\hat{J}_i^y\right\rangle^2 = constant$. This result leads to the emergence of infinite degenerate behaviors associated with the presence of the Goldstone mode.

\section{Phase Diagram}\label{S3}

To facilitate the analytical study, we first investigate the phase diagram of the anisotropic coupled-top model (ACTM) based on the mean-field approach. We neglect quantum fluctuations and correlations between large spins. Within the mean-field framework in the thermodynamic limit $J\rightarrow\infty$, we can construct a trial ground state by a tensor product \citep{MFS1,MFS2} consisting of Bloch coherent states \citep{MFS3}
\begin{eqnarray}
|\psi_{MF}\rangle=|\theta_1, \phi_1\rangle \otimes |\theta_2, \phi_2\rangle=\prod_i\exp\left[\frac{\theta_i}{2}\left(e^{i\phi_i}J_i^-e^{-i\phi_i}J_i^+\right)\right]|J, J\rangle
\label{eq4}
\end{eqnarray}
The reduced expectation values of the large spin operators are represented by the corresponding points on the Bloch sphere as
\begin{eqnarray}
\left(J_i^x,J_i^y,J_i^z\right)=\frac{\left(\left\langle\hat{J}_i^x\right\rangle, \left\langle\hat{J}_i^y\right\rangle, \left\langle\hat{J}_i^z\right\rangle\right)}{J} = \left(\sin \theta_i \cos \phi_i, \sin \theta_i \sin \phi_i, \cos \theta_i\right)
\label{eq5}
\end{eqnarray}
with $(0 \leq \theta_i \leq \pi, 0 \leq \phi_i < 2\pi)$, which can be employed to derive the energy expectation value of the system
\begin{align}
E_{\text{MF}} =& \frac{1}{J} \langle \psi_{MF} | \hat{H}| \psi_{MF} \rangle  \label{EMF} \\
=& -(\cos \theta_1 + \cos \theta_2) + \lambda_x \sin \theta_1 \cos \phi_1 \sin \theta_2 \cos \phi_2 \notag \\
& + \lambda_y \sin \theta_1 \sin \phi_1 \sin \theta_2 \sin \phi_2 \label{EMFeq}
\end{align}
In this manner, the energy of the ground state is determined by values of variational parameters $\theta_i$ and $\phi_i$, which minimize the energy expectation value given in Eq.(\ref{EMF}). Applying the variational principle, we can achieve a set of self-consistent equations:
\begin{subequations}
\begin{align}
&\frac{\partial E_{\text{MF}}}{\partial \theta_1} = \sin \theta_1 + \lambda_x \cos \theta_1 \cos \phi_1 \sin \theta_2 \cos \phi_2 + \lambda_y \cos \theta_1 \sin \phi_1 \sin \theta_2 \sin \phi_2 = 0  \label{VP1} \\
&\frac{\partial E_{\text{MF}}}{\partial \theta_2} =  \sin \theta_2 + \lambda_x \sin \theta_1 \cos \phi_1 \cos \theta_2 \cos \phi_2 + \lambda_y \sin \theta_1 \sin \phi_1 \cos \theta_2 \sin \phi_2 = 0  \label{VP2} \\
&\frac{\partial E_{\text{MF}}}{\partial \phi_1} = -\lambda_x \sin \theta_1 \sin \phi_1 \sin \theta_2 \cos \phi_2 + \lambda_y \sin \theta_1 \cos \phi_1 \sin \theta_2 \sin \phi_2 = 0  \label{VP3} \\
&\frac{\partial E_{\text{MF}}}{\partial \phi_2} = -\lambda_x \sin \theta_1 \cos \phi_1 \sin \theta_2 \sin \phi_2 + \lambda_y \sin \theta_1 \sin \phi_1 \sin \theta_2 \cos \phi_2 = 0 \label{VP4}
\end{align}
\end{subequations}

By carefully analyzing variational equations Eq.(\ref{VP1}-\ref{VP4}), our results demonstrate that the ansatz $\theta_{1}=\theta_2$ is valid throughout the whole regime, whereas solutions for $\phi_i$ exhibit complexity varying across distinct regions. This variation indicates the positional distribution of quantum phases such as PP, FP and AFP.  Then, the quantum phase transitions along the $x-$direction occur at $\lambda_x^{c\mp}$, corresponding to transitions from FP to PP and PP to AFP, respectively. Similar behaviors are observed along the $y-$direction with respect to $\lambda_y^{c\mp}$. The phase transition points are identified as
\begin{align}
\lambda_x^{c\mp}=\mp1, \quad \lambda_y^{c\mp}=\mp1
\label{QPT}
\end{align}

\begin{figure}[ht]
    \centering
    \includegraphics[width=1\linewidth]{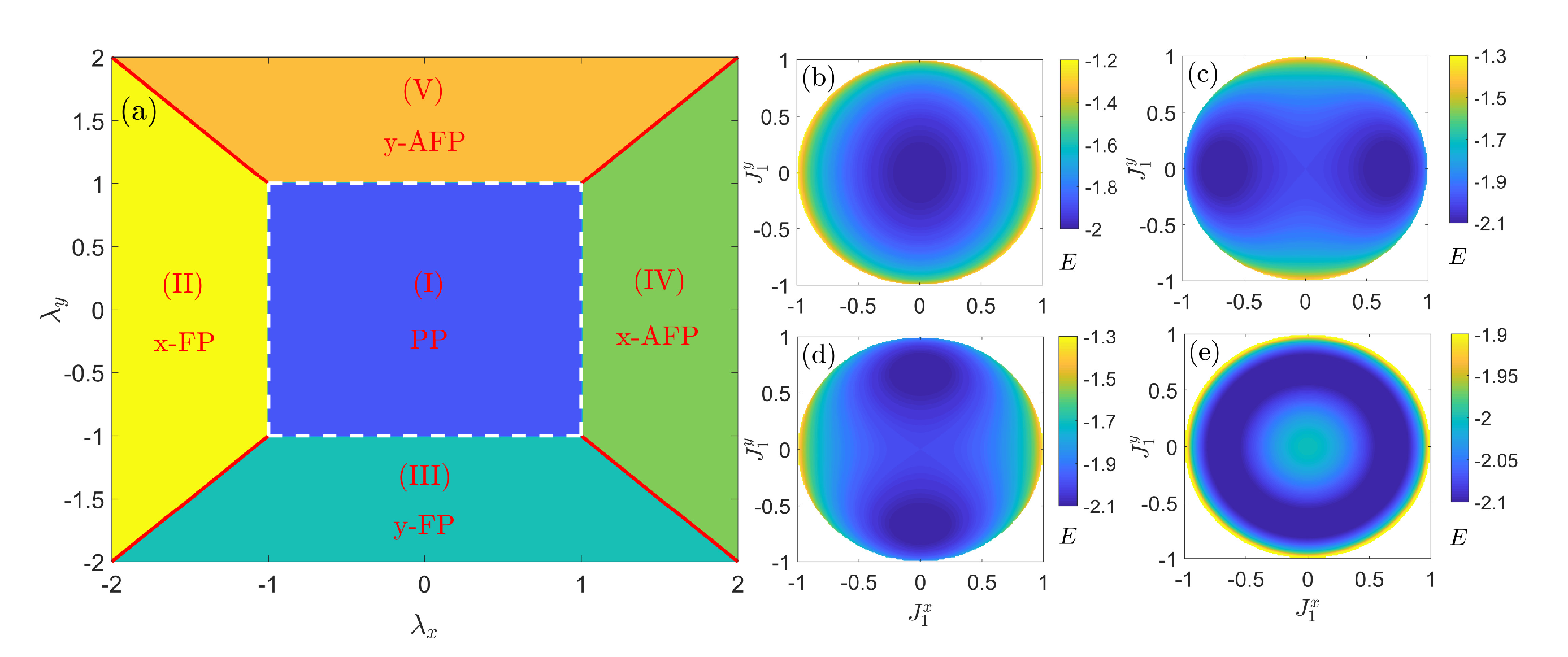}
    \caption{(a) Phase diagram in the
$(\lambda_x,\lambda_y)$ plane. The energy functional $E$ by the mean-field approach for four regions: (b) the PP region for $\lambda_x=0.4$ and $\lambda_y=0.7$ with the energy minimum in the origin, (c) the $x-$AFP region with two minima in the $x-$axis, where the symmetry is broken along the $x-$direction for $\lambda_x=1.4$ and $\lambda_y=0.7$, (d) the $y-$AFP region with two minima in the $y-$axis, where the symmetry is broken along the $y-$direction for $\lambda_x=0.7$ and $\lambda_y=1.4$, (e) the U(1) symmetry-breaking region for $\lambda_x=\lambda_y=1.4$ with a circular valley of degenerate minima.}
\label{PD}
\end{figure}

As shown in Fig.\ref{PD}(a), the white dashed lines separate the disordered phase (PP) from ordered phases (FP and AFP), while the red solid lines distinguish between FP and AFP. Notably, each point along the red lines breaks continuous U(1) symmetry and exhibits an infinite degeneracy. In the following manuscript, we classify the ground states into five distinct phases and provide a detailed analysis of their physical properties. To validate the analytical solutions and symmetry considerations,We numerically minimize the energy functional $E(\theta_1,\phi_1,\theta_2,\phi_2)$ in Eq.(\ref{EMFeq}) with respect to variational parameters $(\theta_2,\phi_2)$ ,  thereby reducing it  to $E(\theta_1,\phi_1)$. To systematically explore the order parameter behaviors, in Fig.\ref{PD}(b-e), we construct the energy function $E$ in terms of the variational $(J_1^x,J_1^y)$, which are determined by $(\theta_1,\phi_1)$.

(i) The symmetry-preserving region, where the disordered paramagnetic phase (PP) dominates, occupies the center of the phase diagram, with $\lambda_d^{c-}<\lambda_d<\lambda_d^{c+}$ $(d=x,y)$. The PP can be characterized by the analytical solutions
\begin{align}
\theta_1=\theta_2=0, \quad E_{\text{MF}}=-2
\label{PPeq}
\end{align}
As evidenced in Fig.\ref{PD}(b), the numerical energy minimum resides at $(J_i^x,J_i^y) = (0,0)$ for $\lambda_x=0.4$ and $\lambda_y=0.7$, confirming the theoretical predictions.

(ii) The symmetry-breaking region along the $ x-$direction, arises when the coupling strength $\lambda_x$ exceeds the critical value while maintaining $|\lambda_x|>|\lambda_y|$. This region can be described by
\begin{align}
\theta_1=\theta_2=\arccos\left( \frac{1}{|\lambda_x|} \right),  \quad E_{\text{MF}}=-\frac{\lambda_x^2 + 1}{|\lambda_x|}
\label{xbs}
\end{align}
Specifically, for $\lambda_x<\lambda_x^{c-}$, the ordered ferromagnetic phase ($x$-FP) dominates with
\begin{align}
\phi_1=\phi_2=0 \   \ or\     \ \pi, \quad J_1^x=J_2^x=\pm\sqrt{1-\frac{1}{\lambda_x^2}}, \quad J_i^y=0
\label{xFP}
\end{align}
For $\lambda_x>\lambda_x^{c+}$, the system stabilizes in the ordered antiferromagnetic phase ($x$-AFP) where
\begin{align}
\phi_i=0, \quad \phi_{i+1}=\pi, \quad J_1^x=-J_2^x=\pm\sqrt{1-\frac{1}{\lambda_x^2}}, \quad J_i^y=0
\label{xAFP}
\end{align}
Notably, the two possible values of $\phi_i$ induce a double degeneracy of the $J_i^x$ in both the FP and AFP. This is consistent with numerical results shown in Fig.\ref{PD}(c), where the energy displays two minima, one with a positive and the other with a negative value, along the $J_1^x$ axis for $\lambda_x=1.4$ and $\lambda_y=0.7$.

(iii) The symmetry-breaking region along the $ y-$direction, occurs when $\lambda_y$ varies and crosses the critical value under the condition $|\lambda_y|>|\lambda_x|$. This regime is governed by:
\begin{align}
\theta_1=\theta_2=\arccos\left( \frac{1}{|\lambda_y|} \right),  \quad E_{\text{MF}}=-\frac{\lambda_y^2 + 1}{|\lambda_y|}
\label{ybs}
\end{align}
clearly, for $\lambda_y<\lambda_y^{c-}$, the ordered ferromagnetic phase ($y$-FP) is dominant with
\begin{align}
\phi_1=\phi_2=\frac{\pi}{2} \   \ or\     \ \frac{3\pi}{2}, \quad J_i^x=0, \quad J_1^y=J_2^y=\pm\sqrt{1-\frac{1}{\lambda_y^2}}
\label{yFP}
\end{align}
For $\lambda_y>\lambda_y^{c+}$, the system enters the ordered antiferromagnetic phase ($y$-AFP) where
\begin{align}
\phi_i=\frac{\pi}{2}, \quad \phi_{i+1}=\frac{3\pi}{2}, \quad J_i^x=0, \quad J_1^y=-J_2^y=\pm\sqrt{1-\frac{1}{\lambda_y^2}}
\label{yAFP}
\end{align}
Analogous to the $x-$direction case, a two-fold degeneracy with dual minima takes place along the $J_1^y$ axis, which is visualized in Fig.\ref{PD}(d) for $\lambda_x=0.7$ and $\lambda_y=1.4$.

(iv) The U(1) symmetry-breaking region, manifests beyond the critical points (indicated by red solid lines in Fig.\ref{PD} for $|\lambda_x|=|\lambda_y|=|\lambda|$).  This region is analytically defined by:
\begin{align}
\theta_1=\theta_2=\arccos\left( \frac{1}{|\lambda|} \right),  \quad E_{\text{MF}}=-\frac{\lambda^2 + 1}{|\lambda|}
\label{u1eq}
\end{align}
Crucially, the infinite degeneracy arises from the continuous freedom of $\phi_i\in\left[0,2\pi\right)$, with the following conditions: $\phi_1-\phi_2=\pi$ for $\lambda_x=\lambda_y=\lambda>0$, $\phi_1+\phi_2=2\pi$ for $\lambda_x=-\lambda_y=\lambda<0$, $\phi_1-\phi_2=0$ for $\lambda_x=\lambda_y=\lambda<0$, and $\phi_1+\phi_2=\pi$ for $\lambda_x=- \lambda_y=\lambda>0$, respectively.
As numerically validated in Fig.\ref{PD}(e) for $\lambda_x=\lambda_y=1.4$, the ground states form a circular manifold of energy minima resembling a Mexican hat profile. All spin polarizations align uniformly in the $\theta-$direction while permitting continuous $\phi_i$ variations, a hallmark of continuous symmetry breaking.

Above all, by adjusting coupling strength $\lambda_x$ and $\lambda_y$, we achieve precise control over the conservation and breaking of $Z_2$ and U(1) symmetries. This capability establishes a promising platform for investigating quantum phase transitions and their associated emergent phenomena, through both theoretical and experimental approaches.

\section{Quantum Criticality}\label{S4}

Previously, we have established the mean-field phase diagram and symmetry properties in Section 3. Now we extend our analysis to quantum criticality induced by higher-order terms and fluctuations beyond the mean-field approximation. To systematically address these effects, first we apply a unitary transformation \citep{TriCT,Rotation} $\hat{\widetilde{H}} = \hat{U}^\dagger \hat{H} \hat{U}$ with
 \begin{align}
 \hat{U} = \exp\left( i\phi_1 \hat{J}_1^z \right)\exp\left( i\phi_2 \hat{J}_2^z \right) \exp\left( i\theta_1 \hat{J}_1^y \right)\exp\left( i\theta_2 \hat{J}_2^y \right)
\end{align}
 where $\phi_i$ and $\theta_i$ are the mean-field solutions from Section \ref{S3}. This unitary transformation rotates spins sequentially around the $z-$axis with angles $\phi_i$ followed by rotations about $y-$axis by angles $\theta_i$. The spin operators in the transformed Hamiltonian $\hat{\widetilde{H}}$ can be expressed as:

\begin{equation}
\begin{pmatrix}
\hat{\widetilde{J}}_{i}^{x} \\
\hat{\widetilde{J}}_{i}^{y} \\
\hat{\widetilde{J}}_{i}^{z}
\end{pmatrix} =
\begin{pmatrix}
\cos\theta_{i}\cos\varphi_{i} & \sin\varphi_{i} & -\sin\theta_{i}\cos\varphi_{i} \\
-\cos\theta_{i}\sin\varphi_{i} & \cos\varphi_{i} & \sin\theta_{i}\sin\varphi_{i} \\
\sin\theta_{i} & 0 & \cos\theta_{i}
\end{pmatrix}
\begin{pmatrix}
\hat{J}_{i}^{x} \\
\hat{J}_{i}^{y} \\
\hat{J}_{i}^{z}
\end{pmatrix}
\label{UT}
\end{equation}

Subsequently, we introduce the mapping from the spin operators to the bosonic creation and annihilation operators via Holstein–Primakoff transformation \citep{HPT}. In the thermodynamic limit $\ (J \to +\infty)$ where $\ J \gg \langle \hat{a}_i^\dagger \hat{a}_i \rangle$, the transformation is expressed as
\begin{subequations}
\begin{align}
\hat{\widetilde{J}}_i^z &= J - \hat{a}_i^\dagger \hat{a}_i \\
\hat{\widetilde{J}}_i^- &= \sqrt{2J - \hat{a}_i^\dagger \hat{a}_i} \hat{a}_i \approx \sqrt{2J}\,\hat{a}_i \\
\hat{\widetilde{J}}_i^+ &= \hat{a}_i^\dagger \sqrt{2J - \hat{a}_i^\dagger \hat{a}_i} \approx \sqrt{2J} \, \hat{a}_i^\dagger
\end{align}
\label{HP}
\end{subequations}
where the bosonic operators satisfy $\left[\hat{a}_i,\hat{a}_j^\dagger\right]=\delta_{ij}$.

By combining Eq.(\ref{UT}) and Eq.(\ref{HP}), we can expand the Hamiltonian $\hat{\widetilde{H}}$ and neglect the terms of order $J^{-s}$ with $s>0$ due to the limit $\ (J \to +\infty)$, yielding:
\begin{equation}
\hat{\widetilde{H}} = J^1 E_{\text{MF}} + J^{\frac{1}{2}} \hat{\widetilde{H}}_{1} + J^0 \hat{\widetilde{H}}_{2}
\label{EXP}
\end{equation}

The first term $E_{\text{MF}}$ represents the ground-state energy obtained from the mean-field approach. The second term shows that $\hat{\widetilde{H}}_{1}=0$ in Eq.(\ref{H1}) as a result of variational equations Eq.(\ref{VP1}-\ref{VP4}). The third term $\hat{\widetilde{H}}_{2}$, determined by mean-field solutions $\theta_i$ and $\phi_i$, corresponds to the quadratic term associated with quantum fluctuations, as detailed in Appendix A. By substituting the values of $\theta_i$ and $\phi_i$ for the $x-$ and $y-$direction symmetry breaking, we can derive the general form of $\hat{\widetilde{H}}_{2}$ in Eq.(\ref{H2}) as
\begin{align}
 \hat{\widetilde{H}}_{2} =&\quad \left(  \cos\theta_1 + |\lambda_{d}| \sin\theta_1 \sin\theta_2 \right) \hat{a}_1^\dagger \hat{a}_1 \notag\\
& + \left(  \cos\theta_2 + |\lambda_{d}| \sin\theta_1 \sin\theta_2 \right) \hat{a}_2^\dagger \hat{a}_2 \notag\\
 &-\frac{1}{2}|\lambda_{d}| \cos\theta_1 \cos\theta_2 \left( \hat{a}_1^\dagger + \hat{a}_1 \right) \left( \hat{a}_2^\dagger + \hat{a}_2 \right) \notag\\
&+\frac{1}{2}f\lambda_{d^{\prime}}\left( \hat{a}_1^\dagger - \hat{a}_1 \right) \left( \hat{a}_2^\dagger - \hat{a}_2 \right)
\label{H2xy}
\end{align}
where $|\lambda_{d}|>|\lambda_{d^{\prime}}|$. It indicates that the index $d=x,y$ designates the principal symmetry-breaking axis, while $d^{\prime}=y,x$ marks the orthogonal direction with preserved symmetry. The discrete parameter $f=\mp1$ serves as a region indicator, corresponding to the regimes $\lambda_{d}<0$ and $\lambda_{d}>0$, respectively. Furthermore, within the symmetry-preserving PP region, Eq.(\ref{H2xy}) is suitable to the form of $\hat{\widetilde{H}}_{2}$ and reduces to $\theta_1=\theta_2=0$.

To systematically diagonalize the quadratic Hamiltonian $\hat{\widetilde{H}}_{2}$, we employ the $\hat{x}-\hat{p}$ representation by constituting the canonical vector $\hat{r}=\left(\hat{x}_1,\hat{x}_2,\hat{p}_1,\hat{p}_2\right)^{T}$ with $\hat{x}_i = (\hat{a}_i^\dagger + \hat{a}_i) / \sqrt{2}$ and $\hat{p}_i = i(\hat{a}_i^\dagger - \hat{a}_i) / \sqrt{2}$. Consequently, $\hat{\widetilde{H}}_{2}$ decouples into the direct sum of two independent subsystems $\hat{\widetilde{H}}_{2} = \hat{H}_x \oplus \hat{H}_p$. Crucially, we illustrated the symplectic transformation\citep{TriCT,Rotation,Wills,SymT} $\hat{S}^T\hat{\Gamma}\hat{S}=\hat{\Gamma}$ governed by Williamson’s\citep{Wills} theorem, which preserves the canonical commutation relations through $\left[\hat{r},\hat{r}^T\right]=i\hat{\Gamma}$, where $\hat{\Gamma}=\begin{pmatrix}
O_2 & I_2 \\
-I_2 & O_2
\end{pmatrix}$ encodes the symplectic structure with $I_2$ and $O_2$ the $2\times2$ identity and null matrices, respectively. Then the quadratic Hamiltonian $\hat{\tilde{H}}_{2}$
 can be exactly diagonalized by the symplectic transformation, manifested as:
\begin{eqnarray}
\hat{\widetilde{H}}_{2} &=& \sum_{i=1,2} \frac{\omega_i}{2} \left( \hat{x}_i^2 + \hat{p}_i^2 - 1 \right) + \tau_{i,i+1} \hat{x}_i \hat{x}_{i+1}+g_{i,i+1} \hat{p}_i \hat{p}_{i+1}  \notag \\
&=& \sum_{i=1,2} \frac{\Delta_i}{2} \left( \hat{x}_i^{\prime 2} + \hat{p}_i^{\prime 2} \right) - \sum_{i=1,2} \frac{\omega_i}{2}
\label{ED}
\end{eqnarray}

In the symmetry-breaking regions where $\theta_1=\theta_2=\arccos\left( \frac{1}{|\lambda_d|} \right)$, above parameters are given by $\omega_1=\omega_2=|\lambda_{d}|$, $\tau_{1,2}=\tau_{2,1}=-\frac{1}{|\lambda_{d}|}$ and $g_{1,2}=g_{2,1}=-f\lambda_{d^{\prime}}$. In contrast, in the symmetry-preserving PP region where $\theta_1=\theta_2=0$, the parameters simplify to $\omega_1=\omega_2=1$, $\tau_{1,2}=\tau_{2,1}=-|\lambda_{d}|$ and $g_{1,2}=g_{2,1}=-f\lambda_{d^{\prime}}$.

Moreover, using the symplectic transformation operator $\hat{S}$, we can construct $4\times4$ covariance matrix $\hat{\sigma}$ \citep{TriCT,Rotation,CoMatrix}, given by $\hat{\sigma} = \frac{1}{2} \left\langle {\left\{(\hat{r} - \langle \hat{r} \rangle) ,  (\hat{r} - \langle \hat{r} \rangle)^T \right\}}\right\rangle=\frac{1}{2}\hat{S}\hat{S}^T$. The quantum fluctuations $(\Delta x_i)^2$ in position space and $(\Delta p_i)^2$ in momentum space are defined by the diagonal elements of the covariance matrix, namely $(\Delta x_i)^2 = \langle \hat{x}_i^2 \rangle - \langle \hat{x}_i \rangle^2 = \sigma_{i,i}$ and $(\Delta p_i)^2 = \langle \hat{p}_i^2 \rangle - \langle \hat{p}_i \rangle^2 = \sigma_{2+i,i}$ for $i=1,2$.

Close to distinct quantum phase transition points $\lambda_{d}^{c-}=-1$ and $\lambda_{d}^{c+}=+1$, we can analytically compute the the excitation energy $\Delta_i$, quantum fluctuations $(\Delta x_i)^2$ and $(\Delta p_i)^2$. Specifically, in the symmetry-breaking regions along $d-$direction ($d=x,y$), where $\lambda_{d}<\lambda_{d}^{c-}$ and $\lambda_{d}>\lambda_{d}^{c+}$, under the condition $|\lambda_{d}>|\lambda_{d^{\prime}}|$ with the the orthogonal direction $(d^{\prime}=y,x)$, these quantities \citep{TriCT,Rotation} are derived as
\begin{subequations}
\begin{align}
\Delta_1 &= \left[( |\lambda_{d}| + \frac{1}{|\lambda_{d}|} ) (|\lambda_{d}| + |\lambda_{d^{\prime}}| )\right]^{1/2}  \\
\Delta_2  &= \left[( |\lambda_{d}| - \frac{1}{|\lambda_{d}|} ) (|\lambda_{d}| - |\lambda_{d^{\prime}}| )\right]^{1/2}  \\
(\Delta x_i)^2 &= \frac{1}{4} \left[ \left(\frac{\lambda_{d}^2-|\lambda_{d}\lambda_{d^{\prime}}|}{\lambda_{d}^2 - 1}\right)^{1/2} + \left(\frac{\lambda_{d}^2+|\lambda_{d}\lambda_{d^{\prime}}|}{\lambda_{d}^2 + 1}\right)^{1/2} \right]            \\
(\Delta p_i)^2 &= \frac{1}{4} \left[ \left(\frac{\lambda_{d}^2 - 1}{\lambda_{d}^2-|\lambda_{d}\lambda_{d^{\prime}}|}\right)^{1/2} + \left(\frac{\lambda_{d}^2 + 1}{\lambda_{d}^2+|\lambda_{d}\lambda_{d^{\prime}}|}\right)^{1/2} \right]
\end{align}
\label{SBq}
\end{subequations}
While in the symmetry-breaking region of the PP for $\lambda_{d}^{c-}<\lambda_{d}<\lambda_{d}^{c+}$ , they are expressed as
\begin{subequations}
\begin{align}
\Delta_1 &= \left[(1 + |\lambda_{d}|)(1 + |\lambda_{d^{\prime}}|)\right]^{1/2}                   \\
\Delta_2 &= \left[(1 - |\lambda_{d}|)(1 - |\lambda_{d^{\prime}}|)\right]^{1/2}                    \\
(\Delta x_i)^2 &= \frac{1}{4} \left[ \left(\frac{1 - |\lambda_{d^{\prime}}|}{1 - |\lambda_{d}|}\right)^{1/2} + \left(\frac{1 + |\lambda_{d^{\prime}}|}{1 + |\lambda_{d}|}\right)^{1/2} \right]                 \\
(\Delta p_i)^2 &= \frac{1}{4} \left[ \left(\frac{1 - |\lambda_{d}|}{1 - |\lambda_{d^{\prime}}|}\right)^{1/2} + \left(\frac{1 + |\lambda_{d}|}{1 + |\lambda_{d^{\prime}}|}\right)^{1/2} \right]
\end{align}
\label{SPq}
\end{subequations}

The entanglement entropy \citep{TriCT,Rotation,VonEntropy} indicates the correlations between two large spins ignored by the mean-field picture. With respect to $\Delta x_i$ and $\Delta p_i$, the entropy can be calculated by
\begin{equation}
S_i = (\Delta x_i \Delta p_i + \frac{1}{2}) \ln \left( \Delta x_i \Delta p_i + \frac{1}{2} \right) - (\Delta x_i \Delta p_i - \frac{1}{2}) \ln \left( \Delta x_i \Delta p_i - \frac{1}{2} \right)
\label{entropy}
\end{equation}

\newpage

\begin{figure}[H]
    \centering
    \includegraphics[width=0.95\linewidth]{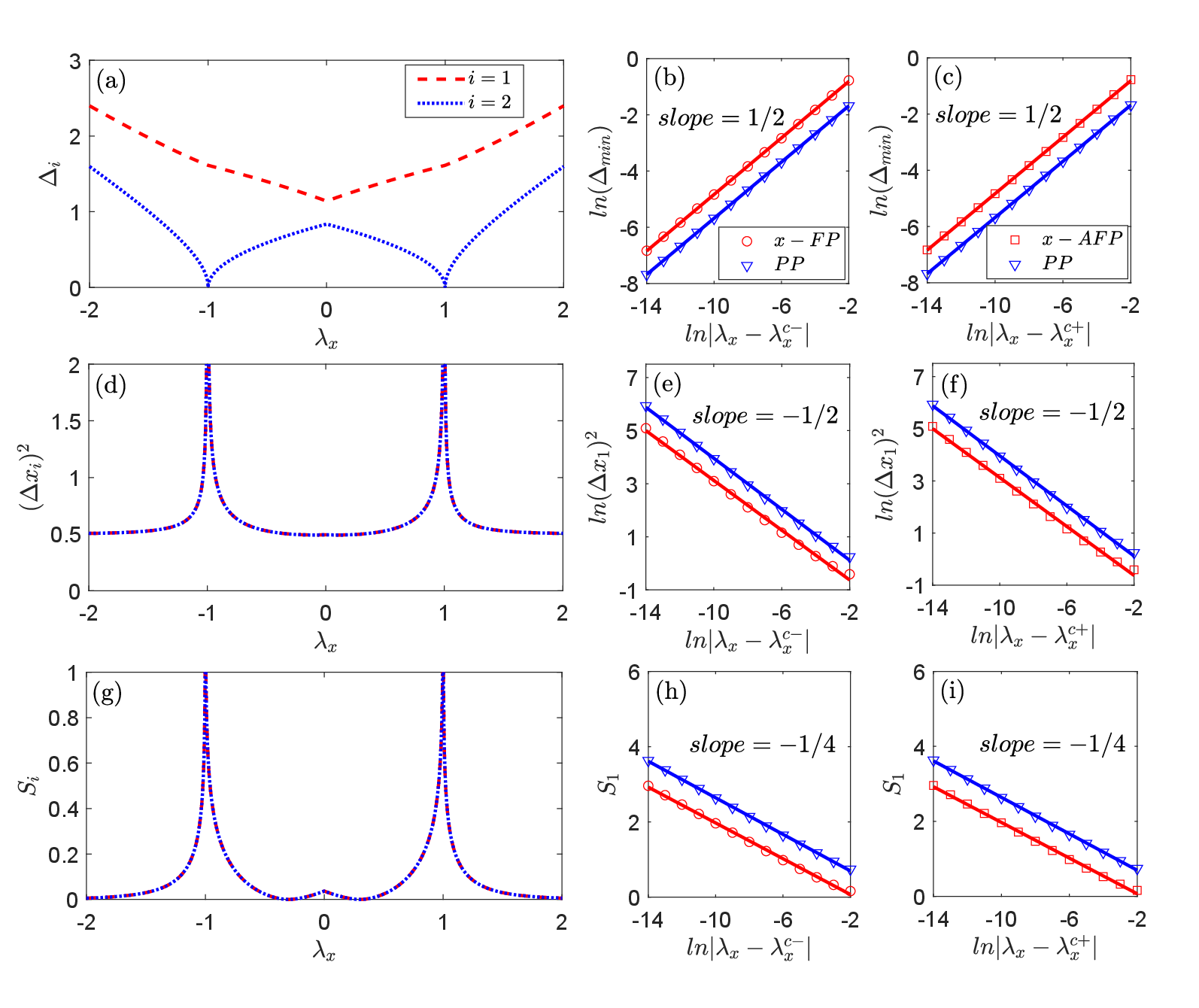}
    \caption{(a) The excitation energy $\Delta_i$, (d) the quantum fluctuation $(\Delta x_i)^2$, (g) the entanglement entropy $S_i$ as a function of the coupling strength $\lambda_x$ for $\lambda_y=0.3$. The red dashed line and blue dotted line correspond to the spin index $i = 1,2$ respectively. The critical points are situated at $\lambda_x^{c-}=-1$ and $\lambda_x^{c+}=+1$, which distinguish the $x-$FP from PP and the PP from $x-$AFP respectively. The ln-ln plot of (b-c) the lowest excitation energy $\Delta_{min}$, (e-f) the quantum fluctuation $(\Delta x_1)^2$ as a function of $|\lambda_x-\lambda_x^{c\mp}|$ near the critical points, respectively. (h-i) show the entanglement entropy $S_1$. The red circles and blue triangles correspond to the analytically exact results for the $x-$FP ($x-$AFP) and the PP respectively, while the solid lines refer to the numerically fitted results. For visibility, the red and blue curves have been shifted to distinguish them.}
    \label{QC}
\end{figure}

\newpage

As shown in Fig.\ref{QC}, we plot the panels for the excitation energy $\Delta_i$, the quantum fluctuation $(\Delta x_i)^2$ and the entanglement entropy $S_i$ as functions of the coupling strength $\lambda_x$ with $\lambda_y=0.5$ held constant. In panel (a), the lowest excitation energy $\Delta_2$, which corresponds to the energy gap of the system, vanishes at the critical points $\lambda_x^{c\mp}=\mp1$. This behavior indicates the second-order quantum phase transitions from the FP to the PP, and from the PP to the AFP, along the $x-$direction. Similarly, the quantum fluctuation $(\Delta x_i)^2$ in panel (d) and the entanglement entropy $S_i$ in panel (g) both exhibit divergences close to the critical points, but tend to stabilize at constant values away from these points. Clearly, these features provide strong evidence for the existence of quantum phase transitions.

All the quantum phase transitions shown in Fig.\ref{QC} are associated with the $Z_2$ symmetry breaking along the $x-$direction. To better understand the universal quantum criticality, it is essential to investigate the critical behaviors of these phase transitions. In panels (b-c),  the ln-ln plots of the lowest excitation energy consistently follow the exponential law $\Delta_{min}\propto|\lambda_{x}-\lambda_{x}^{c\mp}|^{1/2}$ near critical points in four regions: $\lambda_{x}\rightarrow\lambda_{x}^{c-}$ ($x-$FP), $\lambda_{x}^{c-}\rightarrow\lambda_{x}$ (PP), $\lambda_{x}\rightarrow\lambda_{x}^{c+}$ (PP), and $\lambda_{x}^{c-}\rightarrow\lambda_{x}$ ($x-$FP). The divergences of the quantum fluctuation in panels (e-f) and the entanglement entropy (h-i) both exhibit exponential scaling relations in each region, where $(\Delta x_i)^2 \propto |\lambda_{x} - \lambda_{x}^{c\pm}|^{-1/2}$ and $S_i \propto |\lambda_{x} - \lambda_{x}^{c\pm}|^{-1/4}$ respectively. Furthermore, the physical phenomena of symmetry breaking along the $y-$direction are identical to those along the $x-$direction, and are not reiterated in the figure.

Through exact analysis in Eq.(\ref{SBq}-\ref{SPq}) and numerical fitting in Fig.\ref{QC}, we summarize the complete behaviors of the quantum criticality in Tab.\ref{tab1}. These results are consistent with the phenomena observed in the coupled-top model, as well as in the light-matter interaction systems such as the Rabi and Dicke models, both of which exhibit the $Z_2$ symmetry \citep{Rabi3,ARabi3,ADicke}.

\begin{table}[h]
\caption{The excitation energy $\Delta_{min}$, the quantum fluctuation $(\Delta x_i)^2$ and the entanglement entropy $S_i$ near the critical points along $d-$direction for $d=x,y$.\label{tab1}}
\centering
\begin{tabularx}{\textwidth}{lXXX}
\toprule
\hline
\textbf{} & \textbf{FP (\(\lambda<\lambda_{d}^{c-} \))} & \textbf{PP (\(\lambda_{d}^{c-}<\lambda<\lambda_{d}^{c+} \))} & \textbf{AFP (\(\lambda>\lambda_{d}^{c+} \))}\\
\midrule
\hline
\(\Delta_{min}\) & \((\lambda_{d}^{c-}-\lambda_{d})^{1/2}\) & \(|\lambda_{d}-\lambda_{d}^{c\pm}|^{1/2}\) & \((\lambda_{d}-\lambda_{d}^{c+})^{1/2}\)\\
\((\Delta x_i)^2\) & \((\lambda_{d}^{c-}-\lambda_{d})^{-1/2}\) & \(|\lambda_{d}-\lambda_{d}^{c\pm}|^{-1/2}\) & \((\lambda_{d}-\lambda_{d}^{c+})^{-1/2}\)\\
\(S_i\) & \(-1/4\ln(\lambda_{d}^{c-}-\lambda_{d})\) & \(-1/4\ln|\lambda_{d}-\lambda_{d}^{c\pm}|\) & \(-1/4\ln(\lambda_{d}-\lambda_{d}^{c+})\) \\
\bottomrule
\hline
\end{tabularx}
\end{table}

\section{Excitation spectra for U(1) symmetry} \label{S5}

In contrast, by tuning coupling strengths such that $|\lambda_x|=|\lambda_y|=|\lambda|$, the system undergoes the conversation and the breaking of continuous U(1) symmetry, with critical points $\lambda^{\mp}=\mp1$. Nevertheless, these critical points represent the intersections of three phases: PP, FP and AFP. Furthermore, each point in the U(1) symmetry-breaking region exhibits infinite degeneracy, associated with the Goldstone mode.

Using the same transformations as in the previous section, we can derive a similar form of the quadratic term $\hat{\widetilde{H}_2}$ in Eq.(\ref{ED}), determined by mean-field solutions. In the symmetry-breaking regions, the parameters are $\omega_1 =\omega_2=|\lambda|$, $\tau_{1,2}=\tau_{2,1}=-\frac{1}{|\lambda|}$ and $g_{1,2}= g_{2,1}=-k\lambda$, whereas in the symmetry-preserving regions, the parameters become $\omega_1 =\omega_2=1$, $\tau_{1,2}=\tau_{2,1}=-|\lambda|$ and $g_{1,2}= g_{2,1}=-k\lambda$. The discrete parameter $k=\pm1$ serves as an indicator, corresponding to the $\lambda_x=\lambda_y=\lambda$ and $\lambda_x=-\lambda_y=\lambda$, respectively.

\begin{figure}[h]
    \centering
    \includegraphics[width=0.9\linewidth]{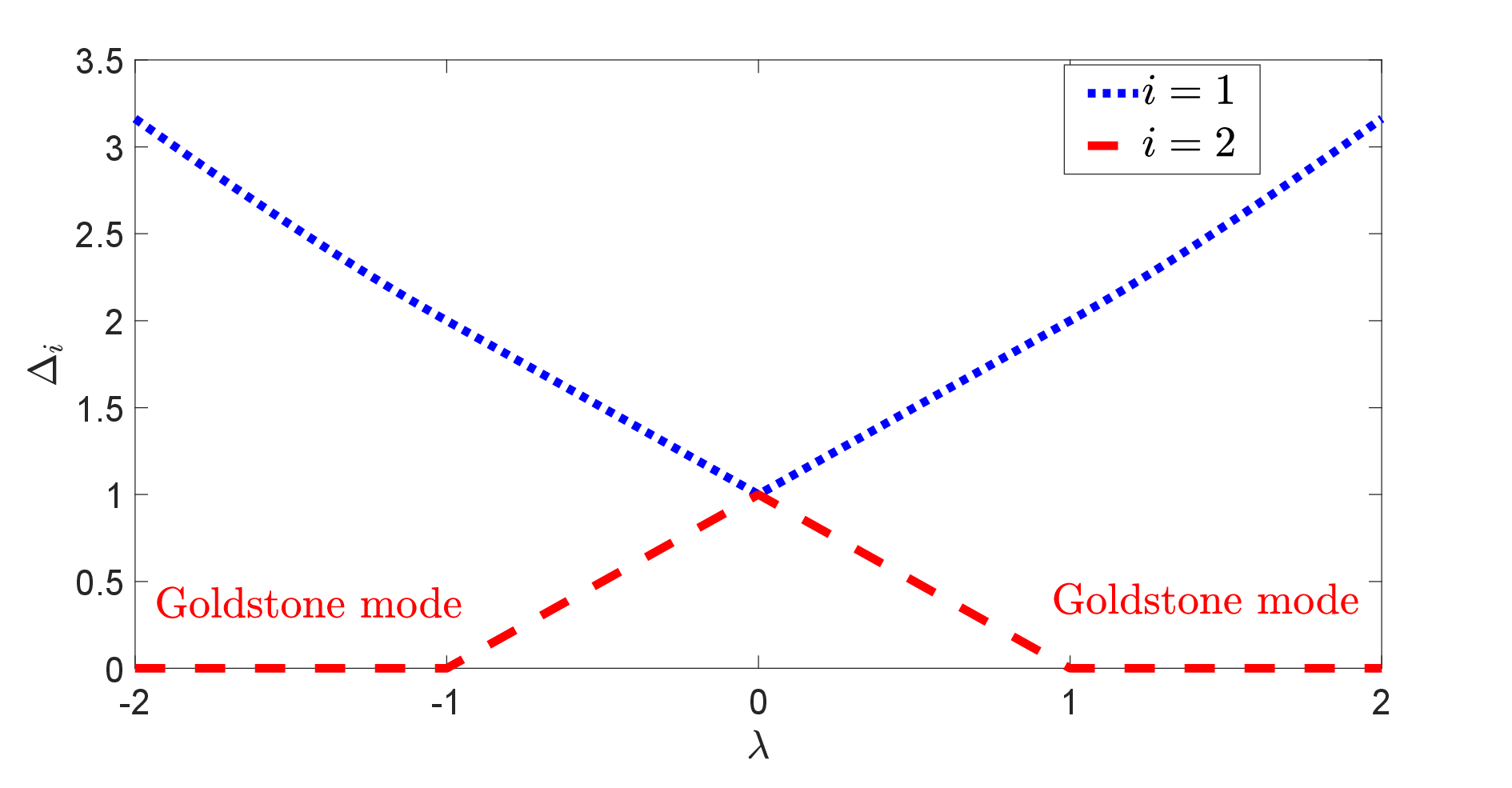}
    \caption{The two branches of excitation energies $\Delta_i$ as functions of $\lambda$ where $\lambda_x=\lambda_y=\lambda$. The energy gap, which is the minimum of two branches, vanishes beyond the critical points, associated with the Goldstone mode.}
\label{ESM}
\end{figure}

Consequently, for $\lambda^{c-}<\lambda<\lambda^{c+}$, the excitation spectra for U(1) symmetry are given by:
\begin{equation}
\Delta_1 = 1 + |\lambda|, \quad \Delta_2 = 1 - |\lambda|
\end{equation}
While in symmetry-breaking regions for $\lambda>\lambda_{c+}$ and $\lambda<\lambda_{c-}$,
\begin{equation}
\Delta_1 = (2 + 2\lambda^{2})^{1/2}, \quad \Delta_2 = 0
\end{equation}

As shown in Fig.\ref{ESM}, two branches of excitation energies represented by blue and red dashed lines, intersect at $\lambda=0$. In contrast to the dispersions shown in Fig.\ref{QC}(a), the energy gap of the lowest branch vanishes beyond the critical points $\lambda^{\mp}=\mp1$, consistent with the infinite degeneracy of the Goldstone mode \citep{JC,ADicke}. Moreover, near the critical points where $\lambda^{c-}\rightarrow\lambda$ and $\lambda\rightarrow\lambda^{c+}$, the energy gap follows a significantly different exponential relation
\begin{equation}
    \Delta_{min}\propto|\lambda_{x}-\lambda_{x\pm}|^{1}
\end{equation}
where the exponent value is $1$ instead of $\frac{1}{2}$ as seen in the $Z_2$ symmetry \citep{JC}.

The vanishing energy gap and the corresponding distinct exponent highlight the uniqueness of the continuous symmetry in contrast to the discrete symmetry. The anisotropic coupled-top model provides an ideal platform for both theoretical and experimental investigations of critical behaviors associated with distinct symmetries.

\section{Conclusions}\label{S6}

In this paper, we introduce a generalized transverse-field Ising model with anisotropic couplings, referred to as the anisotropic coupled-top model. This model describes the interactions between two spin ensembles along both $x-$ and $y-$directions, with distinct coupling strengths $\lambda_x$ and $\lambda_y$. Notably, by tuning these coupling constants, the anisotropic coupling system bridges the gap between discrete $Z_2$ symmetry (for $|\lambda_x|\neq|\lambda_y|$) and continuous $U(1)$ symmetry (for $|\lambda_x|=|\lambda_y|$). The scheme provides an important platform for experimentally controlling the presence and breaking of these symmetries, as well as for investigating the corresponding physical properties.

Using the mean-field approach, we derive the novel phase diagram in the thermodynamic limit $J\rightarrow\infty$, which consists of five phases: the disordered paramagnetic phase (PP), the ordered ferromagnetic phase with symmetry breaking along $x-$ or $y-$direction ($x-$FP or $y-$FP), the ordered antiferromagnetic phase with symmetry breaking along $x-$ or $y-$direction ($x-$AFP or $y-$AFP). When the coupling strengths exceed the critical points $\lambda^{c\mp}=\mp1$ for $|\lambda_x|\neq|\lambda_y|$, the discrete $Z_2$ symmetry breaking occurs in $d-$direction ($d=x,y$), depending on which of $\lambda_x$ and $\lambda_y$ is greater. This results in a double degeneracy of the order parameter along the principal direction. On the other hand, when $|\lambda_x|=|\lambda_y|$, the continuous U(1) symmetry is spontaneously broken beyond the critical points, leading to the emergence of infinite degeneracy associated with the Goldstone mode.

To investigate quantum criticality and fluctuations beyond the mean-field ansatz, we retain and solve the quadratic term in the transformed Hamiltonian using a sequence of transformations. Regarding the $Z_2$ symmetry breaking, we present evidence of the second-order quantum phase transitions, specifically from the FP to the PP and from the PP to the AFP. At the critical points, the excitation energy becomes vanishing, and both quantum fluctuations and entanglement entropy diverge. Furthermore, the analysis of the critical behaviors of these phase transitions supports the universality class of $Z_2$ symmetry. In contrast, when the system undergoes U(1) symmetry breaking, the energy gap vanishes beyond the critical points, yielding a novel exponential relation.

Significantly, our work focuses on the anisotropic couplings between two spin ensembles. It would be valuable to extend this approach to systems involving multiple large spins, such as a single large spin coupled anisotropically to two other spins or more. The rich synergy and competition in these systems provide an ideal platform for both theoretical and experimental investigations. These systems may exhibit complex quantum phase transitions and critical phenomena, thus warranting further exploration.

\textbf{ACKNOWLEDGEMENTS} L.D. is supported by the National Natural Science Foundation of China (NSFC) under Grant No.12305032 and Zhejiang Provincial Natural Science Foundation of China under Grant No.LQ23A050003. Y.-Z.W. is supported by the National Natural Science Foundation of China under Grant No.12105001 and Natural Science Foundation of Anhui Province under Grant No.2108085QA24.

$^{\ast }$ Email:duanlw@gmail.com  \\
$^{\ast }$ Email:wangyanzhi@ahnu.edu.cn

\appendix

\section[\appendixname~\thesection]{Expansion of the  Hamiltonian with HP transformation }
Obviously, the first term is the ground-state energy in the mean-field picture. The second term can be eliminated due to variational equations Eq.(\ref{VP1}-\ref{VP4}), which is detailed as
\begin{eqnarray}
\hat{\tilde{H}}_{1}= &-&\frac{\sqrt{2}}{2} ( \hat{a}_1^\dagger + \hat{a}_1 ) ( \sin\theta_1 + \lambda_x \cos\phi_1 \cos\theta_1 \sin\theta_2 \cos\phi_2     \notag \\
& + &\lambda_y \sin\phi_1 \cos\theta_1 \sin\theta_2 \sin\phi_2)\notag \\
& - &\frac{\sqrt{2}}{2} ( \hat{a}_2^\dagger + \hat{a}_2 ) (  \sin\theta_2 + \lambda_x \cos\phi_1 \sin\theta_1 \cos\theta_2 \cos\phi_2    \notag\\
& +& \lambda_y \sin\phi_1 \sin\theta_1 \cos\theta_2 \sin\phi_2 )
     \notag\\
& + &\frac{\sqrt{2}}{2i} ( \hat{a}_1^\dagger - \hat{a}_1 ) ( -\lambda_x \sin\phi_1 \sin\theta_2 \cos\phi_2 + \lambda_y \cos\phi_1 \sin\theta_2 \sin\phi_2 )        \notag \\
& +& \frac{\sqrt{2}}{2i} ( \hat{a}_2^\dagger - \hat{a}_2 ) ( -\lambda_x \cos\phi_1 \sin\theta_1 \sin\phi_2 + \lambda_y \sin\phi_1 \sin\theta_1 \cos\phi_2 )
 \label{H1}
\end{eqnarray}

The third term $\hat{\tilde{H}}_{2}$ is specifically expressed as
\begin{align}
\hat{\tilde{H}}_{2} = &\quad  \cos\theta_1 \hat{a}_1^\dagger \hat{a}_1 + \cos \theta_2 \hat{a}_2^\dagger \hat{a}_2 \notag \\
&+ \lambda_x \bigg[ \frac{1}{2}\cos \phi_1 \cos \phi_2 \cos \theta_1 \cos \theta_2 (\hat{a}_1^\dagger + \hat{a}_1)(\hat{a}_2^\dagger + \hat{a}_2)  \notag \\
&+ \frac{1}{2i}\cos \phi_1 \sin \phi_2 \cos \theta_1 (\hat{a}_1^\dagger + \hat{a}_1)(\hat{a}_2^\dagger - \hat{a}_2) \notag \\
& + \frac{1}{2i}\sin \phi_1 \cos \theta_2 \cos \phi_2 (\hat{a}_1^\dagger - \hat{a}_1)(\hat{a}_2^\dagger + \hat{a}_2) \notag \\
&-\frac{1}{2}\sin \phi_1 \sin \phi_2 (\hat{a}_1^\dagger - \hat{a}_1)(\hat{a}_2^\dagger - \hat{a}_2)\notag \\
&-\sin \theta_1 \sin \theta_2 \cos \phi_1 \cos \phi_2 (\hat{a}_1^\dagger \hat{a}_1 + \hat{a}_2^\dagger \hat{a}_2) \bigg] \notag \\
&+\lambda_y \bigg[( \frac{1}{2}\sin \phi_1 \sin \phi_2 \cos \theta_1 \cos \theta_2 (\hat{a}_1^\dagger + \hat{a}_1)(\hat{a}_2^\dagger + \hat{a}_2)  \notag \\
&- \frac{1}{2i}\sin \phi_1 \cos \phi_2 \cos \theta_1 (\hat{a}_1^\dagger + \hat{a}_1)(\hat{a}_2^\dagger - \hat{a}_2) \notag \\
&- \frac{1}{2i}\cos \phi_1 \cos \theta_2 \sin \phi_2 (\hat{a}_1^\dagger - \hat{a}_1)(\hat{a}_2^\dagger + \hat{a}_2) \notag \\
&- \frac{1}{2}\cos \phi_1 \cos \phi_2 (\hat{a}_1^\dagger - \hat{a}_1)(\hat{a}_2^\dagger - \hat{a}_2) \notag \\
&- \sin \theta_1 \sin \theta_2 \sin \phi_1 \sin \phi_2 (\hat{a}_1^\dagger \hat{a}_1 + \hat{a}_2^\dagger \hat{a}_2) \bigg]
\label{H2}
\end{align}
determined by mean-field solutions $\theta_i$ and $\phi_i$.


\end{document}